# Title: Locational Factors in the Competition between Credit Unions and Banks after the Great Recession

**Reka Sundaram-Stukel[1] and Steven Deller[2]**

**Author Affiliations**
[1]Dr. Reka Sundaram-Stukel, H. Research Fellow, Department of Economics, University of Wisconsin-Madison.
Corresponding author: rsundara@wisc.edu

[2]Dr. Steven Deller, Professor, Department of Agricultural and Applied Economics, University of Wisconsin-Madison.

**Acknowledgments**
We acknowledge as a group all seminar participants at the Department of Agricultural and Applied Economics. We are grateful for helpful comments from Drs. Barham, Mehta, and Lahkar. We thank our editor Dr. Bond from Cambridge editors for editorial suggestions.

**Funding:** None.
**Conflict of Interest:** None.
**Preprint:** arXiv:2110.07611

**Abstract**

In the aftermath of the Great Recession, the regulatory framework for credit union operations has become a subject of controversy. Competing financial enterprises such as thrifts and other banks have argued that credit unions have received preferential treatment under existing tax and regulatory codes, whereas credit unions complained of undue restrictions on their ability to scale up and increase their scope of operations. Building on previous location models, this analysis focuses on credit union headquarter locations immediately following the 2008 financial crisis. We offer a new perspective for understanding credit union behavior based on central place theory, new industrial organization literature, and credit union location analysis. Our findings indicate that credit unions avoid locating near other lending institutions, instead operating in areas with a low concentration of other banks. This finding provides evidence that credit unions serve niche markets, product differential, and are not a significant source of direct competition for thrifts and other banks.

**Key Points**

1. Credit unions locate within communities where they have a common bond.
2. Credit unions, after the Great Recession, did not show pro-competitive spatial pricing behavior.
3. Credit union use a border location strategy to increase ease in remittance transactions for its membership.
4. Credit union headquarters are not located in areas with high commercial lending because of commercial lending caps.
5. Credit union headquarters locations may be driven by product differentiation, owner-membership organizational form, corporate governance, and regulation.

## 1. Introduction

Independent of the Great Recession, many thrifts and other banks complained about an uneven regulatory environment in their competition with credit unions. Thrifts and other banks maintained that increased credit union consolidations and the relaxation of the common bond criteria in determining credit union membership eligibility undermined the justifications for the special status credit unions enjoyed in tax and regulatory codes. At the same time, credit unions maintained that limits on the types of services they could provide members placed them at an unfair competitive disadvantage. The "feuding" between credit unions and banks had been ongoing and remained unresolved. For example, the American Bankers Association sued the National Credit Union Association for misuse of regulation in 1999. Even though the case was dismissed in 2001, the debate remained heated (Emmons and Schmid 2000, 2003, Feinberg and Meade 2017, and FDIC 2017, DeYoung et al. 2019).

More recently, Maskara and Neymotion (2021) point to results from a Joint Committee of Taxation report that found that the tax-exempt status of credit unions would cost $2.9 billion in terms of lost income tax revenue. Their analysis showed that while credit unions enjoy tax exemption because they serve persons of modest means, this does not mean that they serve underserved populations. The negative economic impacts of the pandemic now begin to recede, the tax-exempt status of credit unions will continue to be hotly debated because neither commercial nor credit unions adequately reach the underserved.

With increased merger activity, expanded product offerings, and significant growth in the size of some individual credit unions, many have begun to resemble traditional banks (Goddard,

McKillop, and Wilson 2002; Feinberg 2008). The relatively faster growth in credit unions over the post–Great Recession period has amplified the argument that the regulatory and tax codes give credit unions an unfair competitive advantage.[1]

Some studies had suggested that banks competing in the same geographic markets did appear to match lending and deposit rates, and therefore credit unions had a pro-competitive effect (Tokle and Tokle 2000, Feinberg 2001, Feinberg and Meade 2017). But does evidence of a pro-competitive effect imply an unequal playing field for thrifts and other banks? Or is it simply the socially optimum level of competition—as intended by regulation? In this study, we focus on parsing out the social versus private incentives of credit union regulation using observed locational patterns.

Organizational differences between credit unions and competing banks make it challenging to verify distortions introduced by regulation. Studies to date have relied on an industrial organization (IO) framework that focuses on market performance—such as market concentration, and more recently efficiency, scale, and scope—to measure competition between credit unions and other banks (Fried et al. 1999, Tokle and Tokle 2000, Wheelock and Wilson 2008, 2011, 2013, Goddard et al. 2002, Feinberg 2001, Bauer et al. 2009; Bauer 2008, Feinberg and Meade 2017, DeYoung et al. 2019, Maskara and Neymotin 2021). Yet the traditional approaches used in this literature, while valid, do not fully account for the monetary and non-monetary aspects of consumer ownership.

---

[1] As not-for-profit cooperative businesses, credit unions are tax exempt and file under 501c14 tax-status. Since they are cooperative businesses, members pay taxes on any residual earnings.

Competition measured from a pure market structure perspective ignores strategic behavior designed to capture market share or rents. In addition, ownership introduces another layer of complexity that limits measurement of market performance in current empirical works. In the credit union context, for instance, switching costs are of singular importance. Depending on the valuation credit union members place on ownership and control rights, it may be a captive market. Another behavioral difference could stem from product differentiation. Credit union members exercise dual roles: as funders of capital and as borrowers of capital. At the minimum, by bundling both saver and borrower services together, consumers-owners may prefer the "one-stop-service" arrangement.

Recent location studies focus on the socioeconomic, solidarity, and inclusivity aspects of credit union location (Pavlovskaya et al. 2020, 2021). Another strand of banking literature investigates whether banks choose their locations strategically to capture market share. Naaman and colleagues (2021) find that compared to credit unions commercial banks are willing to take on more risk. They also find that competition introduces different risk strategies for credit unions. Three studies explicitly study the spatial locations of credit unions and find that they occupy gaps left by commercial banks, although these studies do not provide conclusive evidence that the locational choices are driven by socioeconomic factors (Deller and Sundaram-Stukel 2012, Miller 2015, Mook, Maiorano, and Quarter 2015). This could be because these studies sought to identify a single locational pattern and modeled access to banks based solely on their physical proximity to customers, which assumes open access that credit unions typically do not have. Mook et al. (2015) find that credit union branches are strongly represented in sizable urban communities and are more likely to be located in low-income zip code areas than banks. The data show not only evidence of a credit union niche market but also a tension between social and economic objectives, and that credit unions accommodate themselves to profit norms, which we refer to as market accommodation.

Deller and Sundaram-Stukel (2012) examined credit union location per 10,000 population and argued 1) credit unions do not seem to exhibit “herd” behavior in their location decisions but rather locate based on invoking a common bond criterion, 2) from a strictly spatial location per 10,000 population perspective credit unions are not pro-competitive with retail banks. A pro-competitive effect would result in a credit union forming in the same location as a retail bank, 3) weighted spatial regression showed considerable spatial heterogeneity in credit union locations particularly in areas with high concentration of Hispanic communities. 4) the tax advantage does not elicit pro-competitive behavior. The first, third, and fourth points led to the conclusion that credit unions operate in niche markets.

Building on Deller and Sundaram-Stukel (2012), the present analysis has three contributions. First, this study asks if there is a way to discern the entire nature of competition between banks and credit unions, so that both the known and unexplored sources of friction are revealed using headquarter locations. That is, by examining credit union headquarters data can we elicit an alternative perspective? Second, we place the findings from spatial analysis in broader banking literature. It is important to place this historical reexamination of credit union location in the banking literature because we expect credit unions to play an important role in post-pandemic recovery of financial markets. In the 1970s and 1980s prominent academic journals like the *Journal of Finance* had several key articles on credit unions (Smith, 1984, Murray and White, 1983). Since then, credit unions have received limited attention in the mainstream backing literature and if mentioned it is often in the context of a tax advantage. Admittedly, the insights gleaned from this analysis are more historical than novel. Finally, since credit unions provide access to financial services to people of modest means, and those otherwise underserved by mainstream financial institutions, this paper helps create a more complete record of the state of financial institutions after the Great Recession.

Returning to spatial location patterns we examine the unknown sources of competition between banks and credit unions, by drawing insights from the spatial competition literature and central place theory that tell us that in a competitive environment firms will tend to "herd" together (Shaffer 2004). Based on spatial analysis on credit union headquarters we answer three questions. First, does the spatial pattern of credit union headquarters display strategic behavior to capture rents? Second, are there differences in geographic location patterns that can inform us concerning the nature of competition? New industrial organization theory purports that spatial location can be used to capture rents by 1) bank behavior in its pricing and fragmentation and 2) strategically branching or location to capture rents (Degryse and Ongena, 2012). In this macro-sense headquarter locations shed light on both bank behavior and their strategic objectives. Third, do differences in the socio-economic composition of geographic regions enrich our understanding of competition in the financial sector?

In this study the focus using parallel methodology is shedding light on strategic behavior, spatial pricing, border branching, and rent capture. Spatial analysis has the advantage of generally requiring only zip-code information from the banks, and when juxtaposed with county-level market and socio-demographic information affords a panoramic view of the market and underlying socio-economic factors. With this analysis we also hope to uncover necessary next steps to address the limitations of our approach.

By focusing on the locational decisions of credit unions, our work advances the understanding of competition between credit unions and banks in several ways. We draw attention to the efficacy of using multiple measurement tools to accurately characterize market competition during the 2008 financial crisis. We contribute directly to the discernment of the pro-competitive effect of credit unions in the financial sector by using headquarter location choice analysis. By using U.S. county-level data to approximate local geographic markets, we pay closer attention to socio-economic and demographic features in modeling the spatial competition between credit unions and other banks. We also address an important deficiency in the literature concerning the role of cooperative business structures in the market environment. Finally, this analysis continues to examine credit unions after the financial crisis in 2008, on through 2010. The concerns of that period remain of interest now in a post–COVID-19 world. Our analysis can be immediately applied to understand competition using location choice analysis for other cooperatively organized businesses such as the farm credit system, rural electric companies, and mutual insurance once the U.S. economy returns to normal.

Beyond these introductory comments, the study is composed of four additional sections. First, we review the current thinking about credit unions and banking competition. We then outline our empirical models and estimation methods, which include a simple logit approach to model whether a credit union is present in a county and a zero-inflated Poisson estimator to better capture the number of credit unions present in a county. Using data from U.S. counties drawn from the 2000 U.S. Census, the most current City and County Data Book, and the Bureau of Economic Analysis Regional Economic Information System (BEA-REIS), we model credit union locational patterns. We then discuss empirical results and close the study with a summary of our key findings and a discussion about how these findings may inform the post–COVID-19 competition between credit unions, thrifts, and other banks.

## 2. Literature Review of Credit Union and Bank Competition

Following models from Germany and Canada, the U.S. credit union movement developed in order "to make available to people of small means credit for provident purposes" (Federal Credit Union Act 1934). Credit unions are defined to be member-owned democratic institutions, with an ethos emphasizing self-help and voluntarism especially among the weaker and disadvantaged segments of society. Credit unions pursue a range of social, educational, and developmental objectives. Consequently, credit unions have traditionally been treated differently than banks. Because credit unions are aimed at providing services to markets underserved by traditional banks, certain incentives—such as a favorable tax status— were put in place. In essence, banks did not enter certain markets because they viewed them as unprofitable; to encourage credit unions to enter into these markets, certain incentives needed to be put into place.

Definitionally credit unions can form if they meet the criteria for common bonds of association. The common bonds range from alumni associations, faith-based organizations, homeowners associations and labor unions to scouting groups, parent-teacher organizations within a single school district, and chamber of commerce groups, among others. Exceptions to common bonds include being a Lutheran because there are numerous such churches without a common foundational base. Similarly, being a veteran of the U.S. military is too broadly defined to satisfy the criteria for a common bond. Members of two distinct faith organizations without a common denomination are also not sufficient to form a common bond. Without meeting the common bond criteria and having a sufficient capital base credit unions cannot form as a well-functioning financial institutions under state or federal charters. In this sense, common bonds or association and social capital are pre-requisites for credit union formation.

There are two types of credit unions state and federally chartered credit unions. State-chartered credit unions are managed by state supervisory authority and strive to maintain a local state-specific agenda. A vast majority of state-chartered credit unions are insured by the National Credit Union Insurance Share Fund (NCUSIF) which is supervised by the National Credit Union Association (NCUA). All federally chartered credit unions are governed by NCUA and insured by NCUSIF. Local, state, and government based common bond credit unions declined by 8% they make up the largest group of credit unions. If we look at credit unions with higher asset levels the number of large credit unions declined over the Great Recession, from third quarter 2007 to fourth quarter 2009, and have been steadily declining since. Credit unions with capital less than $10 million in assets declined from 2833 in 2010 to 1049 in 2021—a 63% decline which corresponds to a membership decline of 72% from 2010 to 2021. Noting that this membership base was absorbed by the larger credit unions (see Figure 1 and 2). Smaller credit unions with assets less than $50M increased during the Great Recession. The other concern banks raise is the melting or weakening of common bond criteria. Many of the failing smaller credit unions merged with larger credit unions invoking the multiple common bond criteria. This category where a credit union has one predominant common bond but can incorporate other common bonds has led to a dramatic decrease in credit unions formed under a single common bond.

Two years following the peak of the Great Recession, in 2010, roughly 75% of credit unions had total assets of less than $100 million, while 80% of commercial banks and 85% of savings institutions had assets greater than $100 million. Less than 2% percent of credit unions had assets more than $1 billion (Deller and Sundaram-Stukel 2012). During this time there were 5,036 federally chartered credit unions (FCUs) holding $418 billion in assets and 3,157 state-chartered credit unions (SCCUs) holding $336 billion in assets. While there were a handful of large credit unions, the typical credit union was modest in size, providing services to a small market. By 2010, in response to the Great Recession credit unions in distress merged with healthier credit unions causing a weakening of the "common bonds of association" that dictated membership eligibility.[2] Despite the advent of credit unions offering services increasingly similar to those of ordinary banks (i.e., online banking, branches, and ATM networks), credit unions nonetheless managed to retain their federal tax-exempt status.

This status continues to be a sticking point with commercial banks because six years later, in 2016, the number of credit unions with assets exceeding $1 billion had increased to approximately 272, and by the third quarter of 2021 to 395. An aggregate asset comparison shows that in 2010 credit union assets totaled $849 billion for 7.2% of total commercial assets, and by third quarter 2021 this share had increased to 9.1%. Additionally, if we examine growth rates credit union assets grew by 26.9% from the first quarter of 2020 to the thirds quarter of 2021 and in contrast, commercial bank asset growth has been 17.1% over the same period.[3] Additionally, the trend of smaller credit unions merging into larger credit unions has continued. The number of credit unions with assets of $500M or more has doubled since 2010. The more rapid growth following the Great Recession is likely to strengthen calls for a reexamination of the tax and regulatory status of credit unions in the coming years. This is particularly likely if commercial banks believe that tax exemption confers to credit unions a competitive advantage that outweighs the higher commercial lending volumes enjoyed by commercial banks.

[2] Larger credit unions typically do not have the same common bond criteria as smaller ones.

Pro-competitive effects are traditionally defined as market changes that remove excessive economic profits. Here, the presence of a credit union within a specific geographic market creates competitive pressures on existing banks to improve services. Clearly, credit unions and thrifts or other banks have distinct organizational and regulatory differences. Therefore, in order for banks to validate charges that the competitive environment is unfair, they must demonstrate that credit unions exercised some level of monopoly power, abused the regulatory rules, or that their regulatory status otherwise is outdated or no longer enhances consumer welfare or improves credit access.

Measurement of bank competition is complex and sensitive to research methods and data availability. Looking into the entire spectrum of differences between credit unions and banks, there are significant organizational differences that include governance structure and control rights (Table 1). These are “hidden effects” that can fundamentally alter how banks operate and have serious implications for the competitiveness of financial institutions in the market. However, to date the manner in which these non-monetary differences contribute to competition between banks and credit unions is largely left to speculation.

[3] Some care must be taken comparing percentage increase: over this period credit union assets increased by $431.9 billion while commercial bank assets increase by $3.46 trillion.

While noting the significant regulatory differences between thrifts and other banks and credit unions, we must also recognize that these differences have very little to do with ensuring fair play among banks but rather offer consumer protection. Regulation exists to minimize risk and uncertainty in financial markets and deter costs associated with bailouts of distressed banks (Carletti 2008, Degryse and Ongena 2008).[4] Like banks, credit unions face reserve requirements and mandated insurance coverage to minimize risk and uncertainty. While credit unions benefit from some tax incentives, they face restrictions on scale and scope of operation. At its root the credit union regulatory framework came into being to safeguard working-class Americans by providing access to financial services at affordable prices. A secondary motivation behind this regulation is to increase competition in the financial market.

Credit union regulation has three components, and each is a potential source of conflict. First, credit unions have federal tax-exempt status, under clause 501c14, as not-for-profit cooperative institutions. Second, regulation requires credit unions to form under specified common bond criteria, requiring state- and community-chartered credit unions to reinvest in the communities they serve. Here, the credit union is formed to provide financial services to a group of individuals that are tied through a “common bond of association” such as membership in a labor union.

---

[4] In the event of a crisis, regulation serves to bail out shareholders, directors, and depositors; however, in the current context we view regulation as a mechanism to protect consumers.

Third, most credit unions have a restricted commercial lending portfolio, to less than 12.5% of total assets (NCUA 2009). The first component is a direct subsidy, and most contested by competing banks, whereas the latter two restrict the scale and scope of credit union activities (such as commercial lending and expansion).[5]

The history behind the debate over tax exemption dates to 1999, when the American Bankers Association (ABA) sued the National Credit Union Association (NCUA, 2009) for fraudulent use of regulation. Frame and colleagues (2003) explored the question of whether credit unions had misused their tax-exempt status and found that some credit unions under residential common bond criteria were indeed misusing this status. Credit unions formed under the premise of a residential common bond criterion, however, represent a mere six percent of all credit unions, and these tend to be very small. The ABA case was dismissed in 2001 on the grounds that tax exemption is an integral part of credit unions' non-profit status. The court's ruling, however, has not quelled the debate (Emmons and Schmidt 2003, Hayes 2005). The debate continues, as Maskara and Neymotin (2021) report that the Joint Committee on Taxation found that tax-exempt status resulted in an annual loss of $2.9 billion in 2018 and is projected to result in income tax revenue losses of $3.2 billion in 2020. The current efforts aim to ascertain if credit unions indeed merit this status based on the principles for which they were awarded this exemption (Maskara and Neymotion 2021; FDIC 2017).

[5] Credit unions maintain their reserve requirements with the National Credit Union Administration and purchase deposit insurance from the National Credit Union Share Insurance Fund. Some Massachusetts credit unions purchase private deposit insurance; these are not included in our analysis. Banks maintain their reserve requirements and deposit insurance with the Federal Reserve and the Federal Deposit Insurance Cooperation, respectively.

In an effort to expand the scale and scope of their activities, credit unions have made several appeals to Congress requesting the relaxation of commercial lending caps (Michael 2009; Weiczorek 2010). Most of these were rejected on the grounds that it would compromise the cooperative nature and mission of the institution. More generally, credit unions argue that, irrespective of their size and scope, tax exemption is a function of a not-for-profit cooperative form. To quote CUNA, "$7.5 billion savings to consumers is especially significant when measured against the $1.5 billion in lost federal revenue a year that the government says is represented by the credit union tax exemption." Other than the work of Frame et al. (2003), from a pure regulatory perspective it is not clear that credit unions take advantage of their regulatory status to pursue private gains that should place other banks and thrifts at a disadvantage.

Increased credit union merger activity has added a new dimension to the policy debate between banks and credit unions. Consolidation for banks is said to improve market power or efficiency so as to maximize shareholder value. A non-value-maximizing motive would include: (a) the role of managers pursuing their own interests (where corporate control is weak); and (b) "empire building" to maximize CEO compensation when governance structures are not well defined or not aligned between the board and CEOs. While the level of mergers between credit unions is small compared to mergers among banks, the merger of some larger credit unions coupled with relatively lower losses among credit unions during the financial crisis has fueled the debate. Thrifts and other banks argue that "with all this merger activity and weakening common bond of association, credit unions are just like banks; if only we had the same tax incentives as credit unions—our losses would not be as high" (Michael 2009). In contrast, credit unions counter that "consumer-ownership makes us a good risk, so please increase our ability to compete by relaxing our commercial loan limits" (Michael 2009). In addition, as federal and state regulators are encouraging stronger banks to take over weaker banks, we see the same rationale applied to credit unions. From a policy perspective, the debate appears to have reached a stalemate, and the likelihood of significant policy changes affecting credit unions now appears minimal.

Earlier studies on credit union mergers focused on the price and market concentration effects and found that credit unions did have a pro-competitive effect on loan and deposit rates in local markets (Fienberg 2001; Feinberg and Rahman 2001; Tokle and Tokle 2000; Feinberg and Meade 2017). Placing this work in the broader banking literature, however, it is not clear that we can support the conclusion of a pro-competitive effect in terms of pricing. For instance, Berger and Hannan (1997) and Berger et al. (1999) showed that local markets tended to have near "perfect competition" and smaller banks generally were price takers. Radecki (1998) and Berger and Udell (1998) found that banks with high persistent profit margins were seldom from local markets, and larger banks tended to set deposit and loan rates for local markets. The evidence for linkages between price-setting behavior and local market concentration is weak at best. Davis (2001) found that high reserve requirements limited the expansionary opportunities of credit unions, which tended to grow more slowly than retail banks and other thrifts. Slower growth opportunities in turn dampened the competitive effect of credit unions. Jones and Critchfield (2008) provide a comprehensive review of consolidation activities in the banking industry, and many studies cited in this review show that small business loan pricing continues to have strong price effects in local markets. This is interesting because thrifts and other banks have a larger commercial lending share than credit unions. Thus, from the banking literature in aggregate, it is difficult to conclude that credit unions capture market share/rents as private gains.

Studies find that, post consolidation, there is some evidence to support increased profit and payment system efficiencies, but mixed evidence on cost efficiencies from scale or scope economies (Berger et al. 1997, Berger and Udell 1998, Hughes et al. 2001, 2003; Prager and Hannan 1998).[6] Studies focused specifically on the effect of consolidation on credit unions suggest that efficiency improves for acquired credit unions (Fried et al. 1993; Bauer 2008). Bauer et al. (2009) take this point further to show that the acquiring institution does not show increased efficiency. They also find that most credit union mergers are encouraged by regulators to assist distressed credit unions. This was most evident during the most recent financial crisis. Since credit unions are insured with the National Credit Union Share Insurance Fund (NCUSIF), mergers reduce pressures for insurance to cover fund shortages and losses and help maintain confidence in the credit union system—the "too-big-to-fail" effect. We begin to understand that pure market and pricing approaches are insufficient in isolation to conclude that increased credit union merger activity places thrifts and other banks at risk. The argument that private rent-seeking motivations drive consolidations would have to demonstrate excessive risk-taking or managerial self-interest. While there is evidence to support managerial mischief and increased potential for systematic risk and safety net support among traditional banks, we argue that consumer ownership and a volunteer board makes this less likely for credit unions (Hughes et al. 2001, 2003, Jones and Critchfield 2008). Empirical work on bank consolidation also shows very little benefit to consumers in terms of prices and services (Prager and Hannan 1998, Avery et al. 1999, Avery and Samolyk 2004). Although we would expect that credit unions' organizational form and consumer representation would provide benefits to consumers, especially, in terms of

[6] For a review of the effect of consolidation on the banking industry, see Jones and Critchfield (2008).

ATM networking and branching, as of 2014 there was no empirical work supporting this conjecture. Walker and Smith (2019) did find that since 2012 credit unions acquired 16 thrifts and other banks, and provided better services to their members.

It is clear that banks engage in strategic behavior (hidden aspects) to capture market share and rents, such as product differentiation, switching costs, and geographic location strategies. Kim et al. (2003) focus on vertical differentiation and find that consumers are willing to pay more for banks that offer a higher reputation or superior services. Product differentiation also dictates the degree of substitution between different types of institutions. In the U.S. banking context, Cohen and Mazzeo (2004) study thrifts, single, and multi-market banks and find that competition tends to be fiercest among banks of the same organizational and service type. Another strand of bank literature shows that relational lending or contracts allow banks to extract higher rents from borrowers (Boot and Thakor 2000). Fiercer inter-bank competition pushes banks to offer more relational lending.

All lending in credit unions is relational by definition because members are owners; de facto credit unions serve a captive clientele who are likely to find switching to other financial institutions costly. There is no emphasis in the current literature, however, on the role that switching costs (transaction costs, ATM networks, branching convenience, relational lending) play in understanding the pro-competitive effect of credit unions. We argue that we first need to demonstrate that switching costs are a potential missing link in solving this puzzle. Consumer-level preference data are costly to collect for both banks and credit unions, but location choice is a viable and insightful alternative. Rather than examining bank and credit union competition with respect to pricing, products, and services offered, we therefore recast the question in terms of spatial or geographical competition. In examining the competition between credit unions and other thrifts and banks, we must first observe any existing sources of friction in the markets, such as fragmentation, socio-economic dependencies, and economic rigidities, among other things.

Banks locate or choose branch locations in close proximity to their consumer base for multiple reasons. Location influences spatial pricing as well as the availability of products and services offered. Petersen and Rajan (2002) found that the location of branches and distance to closest competitors are important for both spatial pricing as well as understanding the intensity of competition. Berger et al. (2007) found that most small businesses prefer regional or local banks and that there is a strong correlation between bank reach and national banks. Evidence also suggests that most consumers prefer local or regional banks for long-term loans but rely less on local sources for credit cards. Spatial studies also have shown that banks are able to "redline" certain neighborhoods through strategic branching in predominately higher income white markets, but not in lower income black markets (Avery 1991; Chang et al. 1997). One must keep in mind that from a historical perspective a major motivation for the creation of credit unions is that thrifts and banks were not entering certain geographic markets, specifically, markets that they deemed unprofitable. Unfortunately, while the location of banks has been extensively studied (Amel and Liang 1997, Pilloff 1999, Cohen and Mazzeo 2007) we are aware of only three studies that explicitly models credit unions:   Feinberg (2008), Deller and Sundaram-Stukel (2012), and Maiorano, Mook and Quarter (2016).

Observed “redlining effects” among smaller bank branches and switching costs alert us to fragmentary access or other sources of friction in the financial market. For instance, if traditional retail financial institutions such as banks, and savings and loans institutions tend to geographically cluster in more profitable markets, then we might expect that credit unions would be more prevalent in markets with fewer banks and/or lower profitability. This implies that increased spatial concentration of banks would be met with a lower credit union presence. Furthermore, we would expect to see a higher density of credit unions in poorer or less densely populated areas.[7] Lastly, with a growing Hispanic population in the United States and the widespread presence of credit unions in Central and Latin America, we would expect to see a higher credit union concentration associated with Hispanic neighborhoods. There is considerable evidence documenting increases in remittances to Central and Latin American countries, African countries, and transition economies using credit unions (Grace, 2005; WOCCU 2008, 2010; Ojong, 2014; Kakhkharov, 2020). This type of vertical product differentiation targeting a growing segment of the population is likely be correlated with increased credit unions concentration.

## 3. Empirical Approach, Data, and Economic Estimators

We estimated a family of simple location models using U.S. data for 2,947 counties. Data on the presence of a credit union, our dependent variable, was made available from the University of Wisconsin Center for Cooperatives, the National Credit Union Association call report data, and savings and loans call report data from the Federal Financial Institutions Examination Council for 2009 (Deller et al. 2009).[8] Socio-economic data were drawn from the City and County Data Book 2009. Although we had data on over 8,000 individual credit unions, we aggregated credit unions up to the county unit of analysis to match with the socio-economic (county-level) data. Within this aggregation we had count data on the number of credit unions within each county.[10] We exploited the count nature of credit union data and estimated logit, Poisson, and zero-inflated Poisson (ZIP) models. Deller and Sundaram-Stukel (2012) used zero inflated negative binomial (ZINB) models, and there is some debate as to which approach is better a ZIP or ZINB, we use ZIP because it is asymptotically efficient.

[7] Furthermore, consumers are willing to pay more for banks with better ATM networks than banks without. ATM networks also have an element of horizontal differentiation, whereby consumers pay attention to whether ATMs are conveniently located.

The differences across the financial institutions using three metrics, employee compensation, total assets, and total income for 2009 is illustrated in Figure 3. We chose 2009 as our main year of analysis because the financial crisis peaked in 2008. One striking feature is that larger commercial banks during this period generated income losses. The main takeaway from this graphic is that credit unions are quite small compared to other financial institutions across all three metrics. Total assets of credit unions and other banks are plotted in Figures 2 and 3, respectively. This provides a visual summary of the asset classification by county.

The concentration of credit unions (Figure 4) suggests that there is a spatial clustering in more urban areas. Unfortunately, a simple visual inspection of the location of credit unions as provided in Figure 4 does not definitively inform us about whether there are indeed spatial clusters or if the spatial distribution is purely random. A plotting of the locations of other commercial banks including thrifts (Figure 5), also does not provide evidence of spatial clustering, but does provide evidence of "redlining." Clustering has been discussed within the much broader banking literature as "herding" (Devenow and Welch 1996). Here, the banking literature seeks to better understand why banks tend to act as herds in lending policies (Rötheli 2001), decisions to write down assets (Rajan 1994), and, relevant for our perspective, the decision to open branches and foreign offices within close spatial proximity (Persons and Warther 1997).

[8] http://reic.uwcc.wisc.edu

[10] Note that we only model the headquarters locations of credit unions. Similarly, we have data on the headquarters of over 29,000 thrifts and other banks that we aggregate to the county level. The location of individual branches is not currently available. Because the vast majority of credit unions are too small to support branches, this lack of branch data should not be a serious limitation to this study.

This literature suggests that there are two types of herding: "rational" and "behavioral." The former is consistent with the notion of endogenous (or Porter-style) clusters where the herding is strategic (Porter 2000). There are external economics of scale in terms of locating in close proximity to other financial institutions. The latter, the behavioral approach, exists when banks fall into what can best be described as "groupthink" or more bleakly a "mob mentality." The argument for justifying the behavioral approach centers on information asymmetries and how people (banks) collect, process, and react to information. The rationale is that other banks have access to information that they (the herding banks) do not. Within the rational approach, banks attempt to observe their competitors and learn from their actions. Under the behavioral approach, rather than learn from the actions of competitors, banks blindly follow in an almost knee-jerk mob mentality.

Relevant for the credit union context is that, if banks are indeed clustering according to what might be described as "irrational herding," central place theory tells us that they may be overlooking or avoiding spatial markets that are underserved. These underserved markets may fall below some market threshold for many retail banks, whether this is in terms of population size for remote rural areas or inadequate income and risk profiles (poor, high-risk environments) for some urban markets. This raises the question of whether credit unions are more prevalent in rural and/or poorer communities. As we noted in the literature review we are aware of only three studies that explicitly models the location decisions of credit unions, Feinberg (2008), Deller and Sundaram-Stukel (2012), Maiorano, Mook and Quarter (2016). Unfortunately, a theoretical foundation for Feinberg's (2008) empirical work is lacking. Rather than build a foundation on something like central place theory, Feinberg (2008) appeals to a wide range of bank location studies to justify his empirical specification despite the fact that, as explored below, he is keen to point out that what explains bank location decisions may not apply to credit unions. Maiorano, Mook and Quarter (2016) explore credit unions and bank branch locations within Canada, and because of subtle differences in laws and regulations governing Canadian and U.S. financial institutions, the insights of Maiorano, Mook and Quarter are limited in helping the U.S. situation.

This study is more in line with Deller and Sundaram-Stukel (2012), but rather than modeling the concentration of credit unions, specifically the number of credit unions per capita, in this study we model the absolute number of credit unions in a step process. As such this approach is more consistent with the new industrial organization banking literature (Degryse and Ongena, 2012, Conroy, Deller and Tsvetkova 2017). According to this theory, banking institutions make rent capturing decisions to either affect bank conduct or strategy. The source of rents come from four distinct components: market structure, switching costs, spatial location, and regulation. The analysis in Deller and Sundaram-Stukel (2012) answered the micro-aspects of spatial location and concluded that credit unions did not generate a pro-competitive effect on pricing behavior or market concentration. We explore similar patterns of market fragmentation and branching with headquarters.

If credit unions strategically locate headquarters to capture cross-border rents we should see high concentrations in areas where cross border financial transactions matter. Alternatively, regulation may prevent strategic location. For example, credit unions face commercial lending caps and so it would not make strategic sense for them to locate in commercial districts that focus on high volume of commercial lending. It would make better strategic sense to locate in areas small business seek loans.

We argue that financial institutional behavior drives location decisions, as theoretical spatial models predict (Ali and Greenbaum 1977). For spatial clustering to be the expected outcome, financial institutions have to maximize profits—which most financial institutions do (Hotelling 1929). It is harder to make the case that credit unions are driven by profit maximization because they are member owned, have a cooperative business structure, a cooperative governance structure, are tax exempt, and legally incorporated as 501c1 (federal credit unions) and 501c14 (state-chartered credit unions) entities. Legal incorporation as a cooperative organization prevents them from behaviors typical of traditional financial institution, in that all earned profits minus any member patronage or dividends must be invested back into the credit union.

Disentangling whether credit union location choices reflect strategic spatial clustering versus herd behavior is relatively straightforward. We use techniques common to central place theory and utilize a generalized version of the Getis-Ord Gi* statistics (Ord and Getis. 1995). Anselin (1995) detailed a technique that analyzes spatial patterns across a similar to dissimilar spectrum. Using a set of spatially weighted features (e.g., how geographically close each credit union observation is to nearby or neighboring observations) we compute the Getis-Ord Gi* statistics. Mathematically the Gi* statisticis given by:

$$G_i^* = \frac{\sum_{j=1}^{n} w_{ij}\, x_j - \bar{x} \sum_{j=1}^{n} w_{ij}}{S \sqrt{\frac{n \sum_{j=1}^{n} w_{ij}^2 - \left(\sum_{j=1}^{n} w_{ij}\right)^2}{n-1}}}$$

Where $\bar{x}$ is the mean of the data, or in our case the mean number of credit unions, $w_{ij}$ is a spatial weight defining the spatial interaction (distance) between any two observations, and $S = \sqrt{\frac{\sum_{j=1}^{n} x_j^2}{n} - \bar{x}^2}$ . The $G_i^*$ is a z score where positive values above a statistical significant threshold means a spatial means spatial clustering of high values (i.e., a hot spot), while large negative z scores mean spatial clustering of low values (i.e., a cold spot). In our context, a sufficiently high positive z scores means there is a spatial clustering of credit unions while a sufficiently large negative z scores suggest credit union “deserts.” Areas, or groupings of counties where the z score is neither sufficiently large or small means that there is not statistically meaningful spatial clustering of credit unions.

Considering the results of the the Getis-Ord Gi* statistic (Figure 6), we find remarkable similarities in the identifiable clustering patterns. There appears to be strong hot spot spatial clustering of credit union headquarters in coastal southern California and San Francisco, Seattle, Denver, Chicago, and Detroit, as well as much of Ohio along with the Baltimore to Boston corridor. There is also a clustering along the Gulf coast of Texas and a small group near New

Orleans. There are two cold spot clusters in eastern Kentucky and central Georgia. More important is the lack of concentrations in the vast majority of the U.S. outside of the small geographic concentrations, most of the credit union locations identified in Figure 4 are spatially random. Returning to our central theme, "do credit unions systematically compete with thrifts and retail banks," the lack of spatial concentrations in most of the U.S., other than a handful of metropolitan areas, would suggest the lack of structural competition.

The second pattern that can be inferred from the spatial clustering analysis is the lack of concentration of credit unions in rural areas. The lack of a "rural effect" is relevant to the central theme of this work for two reasons. First, if mergers are the culprits motivating concern about credit unions' supposed regulatory advantage, then the absence of rural clustering or herding patterns makes that argument nebulous. Most significant credit union mergers are urban based. Second, the states in which clustering seems to predominate have a higher immigrant Hispanic population. This means that strategic product differentiation is a likely mechanism for capturing market share rather than pricing behavior (Grace, 2005, WOCCU 2005 2010).

We use four blocks of variables to capture the essence of credit union location patterns in the financial market: competition concentration, economic structure, socio-demographic, and measures of "common bond of association" (Table 2). Concentration variables include bank density per 10,000 persons as well as savings and loans density per 10,000 persons. We hypothesize that credit unions locate in markets that are underserved by thrifts and other banks or that higher concentration of thrifts and other banks in a location will lower the likelihood of credit union presence. In contrast, if credit unions are directly competing with thrifts and other

banks—we would expect to see higher concentrations of thrifts and other banks associated with credit union presence. If credit unions, thrifts, and other banks do not tend to display herding behavior, we can look into the presence of switching costs (product differentiation, specialization, ATM and branch networks). The next steps of analysis from these results would point to barriers to entry, segmentation, and access to financial services.

Socio-demographic variables capture the profile of the markets (counties) that are most likely to attract a credit union. Based on the work of Chang et al. (1997), Avery (1991), and to a lesser extent Feinberg (2008), we would expect to see credit unions headquarters located in poorer, less densely populated counties and those more dominated by minorities. But, at the same time, credit unions require proactive initiatives from local residents to organize, tailor services, and operate. Will counties that are less likely to be attractive to traditional financial institutions have the capacity or social capital to form credit unions? These variables can also inform us about credit unions` strategic behavior in terms of product differentiation and specialization. We also include dummy variables for metropolitan counties as well as adjacent non-metropolitan counties to capture market presence. That is, if credit unions have a significant rural presence, then we would expect to see a negative relationship between each of these variables.

In understanding competition between thrifts and credit unions we must also examine the relationship between different economic structures. For example, are credit unions more or less likely to concentrate in residential or employment-centered counties, or are they attracted to counties with higher or lower concentrations of proprietorships? We include the population to employment ratio along with the population to proprietorship ratio. A county with a higher population to employment ratio is an indicator of a more residential-based county with perhaps a higher share of commuters or retirees. The population to proprietorship ratio is a simple measure of self-employment. We would expect that a higher concentration of self-employed persons is associated with lower levels of common bonds (e.g., potential employee- or union-based credit union membership), thus lowering the likelihood of a credit union being present. These variables fundamentally impact thrifts and other banks, which have higher volumes of commercial lending, with approximately 65% of their lending volume allocated to residential and consumer loans.

We also expand on the base model by controlling for the concentration of organizations of association. Credit unions were traditionally formed with stringent membership criteria based on a "common bond" such as employment, association, religious, or community organization. To test whether one form of organization upon which common bonds can be built is more or less likely to influence the presence or location of a credit union we include: the number of non-agricultural cooperatives; the number of civil and/or social organizations; business associations; professional associations; and number of labor unions—all on a per 10,000 persons basis.

With all else held constant, we expect that higher concentrations of these types of organizations will increase the likelihood of a credit union being present in the county. Also, because credit unions follow the spirit of the cooperative movement, we also include a simple dummy variable for certain types of other non-agricultural cooperatives present in the county. We consider grocery store cooperatives, artisan-focused cooperatives, educational cooperatives, and childcare cooperatives specifically. We limit our attention to these because these specific types of cooperatives tend to be organized around grassroots efforts and reflect the willingness of the community to adopt and use a cooperative business structure. We do not include other types of cooperatives, specifically agricultural and/or utility-focused cooperatives because these types of cooperatives tend to be driven by factors outside of the local community. Their credit needs maybe met by the Farm Credit System. We would expect that credit unions do not have a high presence around business and professional associations because of a restricted commercial lending volume.

We start with the simplest estimation model possible and build on it to gain insights into the questions of interest. The basic estimation addresses a simple question: is a credit union present or not present in the county? This model can be written as $y_i = 1[y_i^* > 0]$, where $y_i^* = X_i\beta_i + e_i$. The binary indicator $y_i$ takes a value of 1 if there is at least one credit union present in the county or 0 otherwise. The logit estimator is derived from the underlying latent variable formulation $y_i^*$ when the error term $e_i$ has a logistic distribution. The vector $x_i$ contains the socio-economic, economic, and common bond characteristics that contribute to the location of the credit union; each co-efficient $\beta_i$ explains the effects of each $x_i$ on the response probability.

Addressing the simple question of whether a credit union is present in the county in a yes-no framework ignores information on the concentration of credit unions. Specifically, we know exactly how many credit unions are located within a county and we can refine our insights into the bank-credit union competition question by taking that additional information into account. In this case, the number of credit unions takes on integer values 0, 1, 2, 3 …. With the conditional mean of $y_i$ given a vector of characteristics $x_i$ denoted by $E(y_i|x_i) = \lambda_i$, where $\lambda_i = \beta' x_i$, then the Poisson distribution is given by:

$$f(y_i = y|x = \frac{exp[-\lambda_i](\lambda_i^y)}{y!}), y = 0,1,2,3, \ldots \quad (1)$$

The Poisson distribution imposes restrictions on the conditional moments; most commonly, the variance equals the mean: $Var(y_i|x_i = E(y_i|x_i) = \lambda_i$ . While the estimation of this model is fairly straightforward, the variance structure is often violated. Specifically, when the variance is greater (less) than the mean, there is over-dispersion (under-dispersion) relative to the Poisson distributional case. Over-dispersion is a potential source of concern because it is quite possible to have counties without any credit unions. To circumvent this, we envisage a two-step process. First, owing to certain socio-economic or income characteristics, some counties may not attract credit unions. These counties will always show a zero number of credit unions independent of the underlying data-generating process. In other counties, in contrast, the number of credit unions follows a Poisson process, but nonetheless may show no credit union presence due to the data-generating process.

A natural way to model this was suggested by Lambert (1992) and Papoulis (1984) by putting a point mass at 0, because those cases where counties have no credit unions are also of interest.[11] That is, with probability $p$ we sample from a degenerate distribution at 0, and with probability 1-$p$ we sample from a Poisson $\lambda_i$ distribution. This model is commonly referred to as the zero-inflated Poisson (ZIP) model. Explicitly given the parametric distribution $\pi(y)$ on integer values $y = 0,1,2,3, \ldots$ ,we can write the associated zero-inflated distribution as:

$$P(y_i|p,0) = p + (1-p)\pi(y),\ P(y_i|p,0) = (1-p)\pi(y), y > 0 \quad (2)$$

[11] For notational consistency we follow Papoulis (1984).

Where $p$ is the probability, we use a logistic specification on a latent indicator variable $z$; and we define a joint distribution to estimate equation (2) as follows:

$$P(y_i = 0, z = 1|p) = p,\ P(y_i = y, z = 0|p) = (1-p)\pi(y) \quad (3)$$

Then $z$ is a Bernoulli random variable with a probability of success of $p$. Then given a sample size $n$ with $y_i$ given $p_i$ distributed as in equation (2), the marginal or observed likelihood can be written as:

$$L(p, y) = \prod_{i=1}^{n} [p_i(1(y_i = 0) + (1 - p_i)\pi(y_i)] \quad (4)$$

where $\pi(y)$ is $Poisson(\lambda)$ and leads to the zero-inflated Poisson model, $ZIP(p, \lambda)$.The $ZIP(p, \lambda)$ model is over-dispersed relative to the $Poisson(\lambda)$: that is, $E(y|p, \lambda) = (1-p)\lambda < \lambda$ and $Var(y|p, \lambda) = (1-p)\lambda(1+p\lambda) > E(y|p, \lambda)$. The spatial association in observed counts of credit unions is explained by each $\lambda_i$. Indeed, this would be the case if each $z_i$ were observed. But we argue that the counties showing no credit unions are unobserved, so introducing a spatial effect for each observation would make the estimation unstable. In our sample, over half (50.5%) of the counties do not have credit unions and we find very little evidence of spatial clustering outside of a small handful of urban areas (Figures 6 and 7), so we keep the estimation aspatial.

## 4. Estimation Results

Recall we are interested in understanding what the spatial location of credit unions can tell us about the nature of competition in the market. The results of the logit, Poisson, and ZIP modeling are presented in Table 3. In general, the results across the three estimators are consistent, with the zero-inflated Poisson (ZIP) estimation providing the strongest set of results. We did find a substantial number of cases where variables are statistically significant in one model but not in another, with the ZIP model yielding the most statistically significant variables. More importantly, there are few cases where the same variables were statistically significant but with different directional relationships (signs) across estimators. This lends credence to our view that the ZIP model is the right specification in this context. Before turning attention to the principal question of interest, consider the results of some of our socio-demographic variables.

The racial composition of counties did influence the location decisions of credit unions. All three variables—African-American, Hispanic, and foreign born—appeared to have a positive influence on the presence of a credit union within a county. Of these, the percentage of Hispanic population was statistically significant across the three models (logit, Poisson, and ZIP).

The higher concentration of credit unions in U.S. counties with large concentrations of Hispanics is not surprising. Latin and Central American countries have approximately 1,800 credit unions with $38 billion in total assets. Because credit unions offer remittance services to their consumers they are particularly attractive to Hispanics. The closest alternative to credit unions for remittances for working-class and migrant workers is Western Union, but with high associated transaction fees. We would therefore expect that credit unions, by making both accounts and availability of ATM services easily available to the recipients of remittances, would have high market concentration in Hispanic communities. A higher share of the population that is African American and the percent of the population foreign born, while not significant in the logit model, are significant in the Poisson and ZIP models. Here, a higher concentration of both is associated with a higher concentration of credit unions, all else held constant. These results are consistent with prior expectations because these are generally market populations that banks and

other thrifts may underserve. Educational attainment modeled by percentage of population over 25 with a bachelor`s degree is positive and statistically significant.

The poverty rate and unemployment rate have a significant negative influence credit on union concentration in the ZIP model. The potential negative relationship of credit unions with poorer areas is potentially a source of concern. Thrifts and other banks have argued that they serve a higher share of the underserved population than credit unions, however, evidence does not seem to support this. While results from the present analysis are not adequate to explain the percentage of the underserved market captured by credit unions versus their closest competitors, it does point to the need for further work. We could argue that credit unions require a minimum level of social capital within the community to support their formation, and a generally accepted hypothesis is that social capital and poverty are inversely related. It is equally likely that our aggregate measure of poverty is not sufficiently refined to fully address the hypothesis as currently specified.

The percentage change in the number of households, a socio-demographic variable that also reflects to a certain extent economic growth, has a statistically significant negative impact on credit union concentration. This is as expected, because traditional retail financial institutions such as thrifts and other banks are drawn to more profitable markets that are experiencing economic growth, which would be reflected, among other things, in higher levels of household formation. If credit unions avoid locating near traditional retail banks, then we would expect counties that are experiencing economic growth would see lower concentrations of credit unions. Owner occupancy rates also negatively influence concentrations in the ZIP model. Higher values of the population-employment ratio are associated with lower concentrations of credit unions. This suggests that credit unions are more likely to be located in counties that can be described as employment hubs as opposed to residential or bedroom communities. Given the “common bond of association” requirement for credit union formation, this result may be interpreted as indirectly capturing credit unions, which are based on employment opportunities. Conversely, the population-proprietor ratio is positive and significant in all models.

Now we consider the principal question of interest: are credit unions located in close spatial proximity to banks? Our evidence suggests the contrary—a higher concentration of banks, measured by banks per 10,000 persons, is associated with a lower concentration of credit union headquarters. This means that credit unions do not exhibit the same herding behavior that is characteristic of the banking industry. In fact, the results indicate that credit unions avoid locating where competing banking services are available. The results do indicate, however, that thrifts (referred to as savings and loans) and credit unions are spatially located in close proximity, which suggests herding behavior between the two. Closer inspection, however, reveals that credit unions serve a fundamentally different segment of the market. Even in presence of spatial price equalization between thrifts and credit unions, ownership and governance characteristics of credit unions may create strong incentives for a captive consumer base (Degreyse and Ongena, 2008). That is, as the owners of the credit unions, the consumer base is locked into a rent-shielding relationship. Thus, to compare competition between thrifts and credit unions we would need to take into account barriers to entry, vertical and horizontal differentiation, and switching costs, rather than spatial pricing (Degreyse and Ongena, 2008).

Turning to factors that lend support to the presence of barriers to entry (e.g., common bond variables), community-based grassroots cooperatives do appear to have a positive influence on the concentration of credit unions and are statistically significant across all three models. In general, there are strong and consistent results on the five measures of “common bond of association.”

The larger the number of civil-social organizations per 10,000 persons, the higher the concentration of credit unions. This result is consistent across all three estimators provided in Table 3. We suggest that this measure not only captures one potential source of common association but is also a measure of social capital. The number of business associations per 10,000 population, however, has a negative influence on the concentration of credit unions. This, again, is an expected result because credit union business lending volume is restricted to 12.5% of total assets. Thus, communities (counties) that have a high concentration of business associations are more likely to be attractive to retail banks, which in turn repel the formation of credit unions. The concentration of professional organizations does not appear to influence the creation of credit unions. The strongest result appears for the concentration of labor unions, which has a positive influence on credit union concentration. The formation of credit unions as a service to their membership has been a common practice for labor unions. The strong ties to a common bond show that credit unions do indeed serve a fundamentally different consumer base, emphasizing once again the rigidities in the market in form of product differentiation, switching costs, and barriers to entry.

But when we focus on the key variables of interest—such as population density, the metro and non-metro adjacent dummies, the concentration of thrifts and other banks, and finally the measures of "common bonds of association"—none of these demonstrate any spatial heterogeneity with respect to credit unions. Variables that do exhibit spatial heterogeneity in terms of the presence of credit unions include percent of the population that is African American, Hispanic, and foreign born, the poverty rate, change in number of households, percentage of houses owner occupied, the unemployment rate, and the population-proprietorship ratio if we lower the confidence level.

Combining the results of this study with Deller and Sundaram-Stukel (2012), we begin to see that to really address competition between thrifts, other banks, and credit unions we must examine the role of switching costs, value of ownership, product differentiation, and barriers to entry.

## 5. Discussion

Based on results presented here and in other studies, it becomes clear that comparing the subtleties of competition between two fundamentally different types of financial institutions is fraught with imprecision and cannot encompass the full spectrum of organizational variation across business types. Our principal contribution to the literature is rather to draw attention to the organizational structure of cooperatives and how they influence competition among other firms offering similar services. A few consistent patterns emerge. First, credit unions served fundamentally different segments of the market than thrifts or retail banks. On the whole, there is weak evidence of herding between credit unions, thrifts, and other banks, as the results are not statistically significant across all models. Our spatial analysis models point to spatial clustering of credit union headquarters in urban areas along coastal southern California, San Francisco, Seattle, Denver, Chicago, Detroit, Ohio, Baltimore to Boston corridor. Aside from these noted areas there is lack of concentrations of credit union headquarters and most of the credit union locations identified in Figure 4 are spatially random.

There has been much speculation about the pro-competitive effect of credit unions on the interest rates and fees offered by retail banks. We set out to explore whether credit union regulation has created an unfavorable competitive environment for thrifts and other banks. It appears from this analysis that such competition is infrequent, as credit unions tend to avoid areas that have a higher concentration of thrifts or other banks. There is weak evidence of locating in close proximity to thrifts.

Second, we find that common bonds of membership appear to be the source of segmentation. That is, credit unions exhibit a strong positive concentration in areas with labor unions, the presence of other cooperative organizations, and civil and social organizations. They do not, however, show a strong market concentration in areas where professional and business associations are numerous, thus maintaining their ties closely to serving working-class Americans. These results are not surprising, because most thrifts and community banks are involved in a larger volume of commercial lending and most small-business lending tends to occur through relational contracts. Third, home ownership and the level of business proprietorship are negatively associated with credit union concentration, reiterating the finding that credit unions have a client base that is rooted in service rather than business activity. Once again, thrifts and other banks are required to hold 65% percent of their lending volume in residential and consumer loans, so these results are consistent with market segmentation. Fourth, poverty and unemployment do not seem to increase credit union concentration, but racial and ethnic affiliation—particularly the presence of Hispanic communities—show a higher credit union concentration. These results are aligned with the objectives of the World Council of Credit Unions to ensure the easier flow of funds via ATM-networking, wire transfers, and online banking to Central and Latin American credit unions (Grace, 2005; WOCCU, 2005, 2008). Fifth, credit union headquarters are not typically located in the poorest neighborhoods. Credit unions form under a common bond, so they need to have an initial wealth base to support banking (Grace, 2005;  Kakhkharov and Rohde, 2020).

To conclude, if spatial patterns of credit union location can be viewed as an accurate measure of "yardstick competition," then credit unions occupy a distinct market niche. It is not clear if at the margin the members of these credit unions would switch allegiance to regular banks if the credit unions did not offer competitive rates. Credit union headquarter locations matter more than branch locations because they paint an accurate picture of the strategic motivations behind location choice. For example, the UW credit union serves the University of Wisconsin, so it would not make sense to locate in Chicago University campus. Similarly, it would not make sense the Naval credit union to locate in the financial district in lower Manhattan. Furthermore, our results show that to accurately discern the private versus social pro-competitive effects of credit union regulation we must pay careful attention to organizational form, switching costs, product differentiation, barriers to entry, the network of auxiliary services, and cross-border activity.

Tying these findings back to new industrial organization theory, our results show that credit union headquarters are strategically located in areas where strong common bonds exist or where there are opportunities to offer services not offered by other banks. They are strategically not located in commercial and business areas like other banks because of regulation (e.g. caps on commercial lending). As consumer-owners of credit unions, members are locked into a rent-shielding relationship, which keeps credit unions tied to their community roots and enforces community bonds and relational lending. Another factor influencing location choice is bank pricing behavior. Although Feinberg (2001) found evidence for spatial price discrimination, we do not find the same evidence to support spatial price discrimination post Great recession.

In the credit union context, bank organization and corporate governance play a significant role in cross national border banking where remittances are common and the culture of cooperative banking is strong; this includes Latin American, African, and European countries. In the post-pandemic recovery period, this culture of cooperative banking may result in increased branching or headquartering in areas with high concentrations of remittance transactions. Banking borders affect location through regulation. We would expect credit unions to opt out of locating in areas with high business or commercial activity because of the regulations that limit commercial lending. It would make strategic sense to locate in areas where civil and societal organizations or trade and labor unions are strong.

This said, some limitations and directions for future empirical work are important, especially for understanding the formation of credit unions and mergers in a post–COVID-19 financial world. Unlike the Great recession where the financial collapse and bank mergers had their roots in subprime mortgage lending, the COVID-19 based economic downturn has been heavily supported by financial buffers and sound monetary policy (IMF, 2020, 2021,2022). As economies recover from the pandemic, credit unions could strategically increase the number of credit union branches or headquarters in the U.S. in order to support remittances to both transition economies and low middle-income economies. In this context, the formation of new credit unions under the common bond criterion would be based on belonging to community organizations within these ethnic groups. Our results, for the period 2008-2010, are conservative since we model only headquarters locations. Examining the spatial patterns of branch and ATM networks for both banks and credit unions would push our understanding further. Given that most merger activity in our study occurred in urban locations, separating the analysis of performance by urban and rural effects is equally important. Using direct measures—such as switching costs, product differentiation, and barriers to entry among others—appears to be a logical direction for future empirical work as well. But developing a theoretical framework that explicitly models cooperative business performance is the most important and immediate task.

## 6. Conclusion

This paper sought to revisit location decisions for credit union headquarters after the Great Recession in 2009 to provide historical context to credit union mergers and assess the pro-competitive nature of credit unions with other banks. While we did not find evidence of pro-competitive behavior with other banks, new industrial organizational theory offered some insights into motivations for location choice using bank strategy and bank behavior frameworks.

Credit unions strategically locate headquarters in neighborhoods with high concentrations of social and civil associations because of common bond criterion. We found remittances may also drive bank border headquarter location choices. Because of the dual owner-consumer relationship, transaction costs, switching costs, service bundling, and product differentiation, credit union members are locked into relational banking. Unlike merger behavior after the Great Recession we hypothesize that increased credit union headquartering and branching may be the response to the post-COVID-19 recovery. We further acknowledge that banks could strategically branch in reaction to credit union locations and this merits its further investigation.

Table 1: Differences between Credit Unions and Other Banks in terms of Size, Organization, Objectives, Regulation, and Measurement of Competition

| Sources of Differences | Thrifts and Commercial Banks | Credit Unions |
|---|---|---|
| **Size and Scale** | | |
| | 8,000 communities and thrifts<br>Approximately 80% have assets higher than $1 billion, accounting for 23% of total assets of U.S. banking industry. | 8,193 credit unions<br>< 2% have assets higher than $1 billion. |
| **Organizational** | | |
| 1 | Investor owned | Consumer owned |
| 2 | Board members are hired from community members and small businesses within the community. | Board members are elected from consumer-owners of the credit union. |
| 3 | Board members are compensated and hold a high equity state in the bank, | Board members are uncompensated because of by-laws. Board members do receive educational training and per diem expenses for board-related activities. |
| 4 | Consumers of thrifts and commercial banks do not have a voice in the way the bank operates. | Consumers have control rights and voteon both consumer and producer decisions. |
| **Objectives and Profits** | | |
| 1 | Maximize rents that is producer surplus | Maximize rents that is producer and consumer surplus. In the case of credit unions consumers are producers. Credit union members have a dual relationship with the credit union. |
| 2 | Profits are distributed to the shareholders and potentially also reinvested in the bank. | Profits can be redistributed as lower loan rates, higher deposit rates, dividends, or held back as retained earnings for credit union growth. |
| **Regulation** | | |
| 1 | Taxed as for-profit C or S-corporation | Tax exemption as not-for-profit 501c12. Paid taxes are property taxes, payroll taxes, and income taxes on earned interest and dividends. |
| 2 | No common bond requirement | Common bond requirement and community reinvestment requirement (state-chartered credit unions) |
| 3 | Required to maintain a lending portfolio of 65% in housing or consumer markets and 23% in commercial or small business lending | Commercial lending cap of 12.5% of total assets |
| 4 | FDIC reserve requirements | NCUSIF reserve requirements, which are typically higher than what FDIC recommends |
| **Measurement of Competition** | | |
| | Scale and scope economies, efficiency, Xefficiency, productivity, and risk management. | Scale and scope economies, efficiency, Xefficiency, productivity, and risk management |

Sources 2010: National Credit Union Association (NCUA), American Banking Association, and Independent Community Bankers of America

Table 2: Variables Used in the Credit Union Location Model

| Variables | Potential Inference for Credit Unions |
|---|---|
| **Market Concentration** | |
| Population density | Access and product differentiation |
| Metro County | Access and entry |
| Non-metro adjacent | Access and entry |
| Number banks for 10K population | Product differentiation, network, and specialization |
| Number savings and loans for 10 K population | Product differentiation, network, and specialization |
| **Economic Structure** | |
| Unemployment rate | Access |
| Population to employment ratio | Access and specialization |
| Population to proprietorship ratio | Access and specialization |
| Poverty rate | Access |
| **Socio-demographic Variables** | |
| Percent of population African American | Access and specialization |
| Percent of population Hispanic | Access, product differentiation, network, and specialization |
| Percent of population over age 25 with bachelor's degree | Access, product differentiation, and network |
| Percent of population foreign born | Network and specialization |
| Percent of change in number of households 2000-2005 | Network |
| Percent of houses owner occupied | Specialization and network |
| **Common Bonds of Association** | |
| Non-agricultural cooperatives present | Specialization |
| Number of civil-social organizations per 10K population | Network, product differentiation, and specialization |
| Number of business associations per 10K population | Network, product differentiation, and specialization |
| Number of professional associations per 10K population | Network, product differentiation, and specialization |
| Number of labor unions per 10K populations | Specialization and product differentiation |

Source: Census county data and IMPLAN database

Table 3: Estimates for Credit Union Location Patterns: Dependent Variable Number Credit Unions

| | Logit<br>Yes=1, No=0 | Poisson | Zero Inflated Poisson |
|---|---|---|---|
| Intercept | 2.1478 ** | 1.8200 *** | 4.1875 *** |
| | (0.0200) | (0.0001) | (0.0001) |
| **Market Concentration** | | | |
| Metro County | 0.6958 *** | 1.1481 *** | N.A. |
| | (0.0001) | (0.0001) | |
| Non-Metro Adjacent | 0.0208 | 0.0864 | -0.5655 *** |
| | (0.8550) | (0.1070) | (0.0001) |
| Population Density | 0.0050 *** | -0.0001 * | -0.0001 *** |
| | (0.0001) | (0.0510) | (0.0001) |
| Number of Banks per 10K Population | -0.2035 *** | -0.1475 *** | -0.3145 *** |
| | (0.0001) | (0.0001) | (0.0001) |
| Number of Savings and Loans per 10K Population | 0.0543 | 0.2863 *** | 0.3729 *** |
| | (0.4860) | (0.0001) | (0.0001) |
| **Socio-Demographic** | | | |
| Percent of Population African American | -0.0022 | 0.0115 *** | 0.0140 *** |
| | (0.5930) | (0.0001) | (0.0001) |
| Percent of Population Hispanic | 0.0166 ** | 0.0047 ** | 0.0084 *** |
| | (0.0040) | (0.0020) | (0.0001) |
| Percent of Population over Age 25 with a Bachelor's Degree | 0.0267 ** | 0.0122 *** | 0.0151 *** |
| | (0.0100) | (0.0001) | (0.0001) |
| Percent of the Population Foreign Born | -0.0156 | 0.0331 *** | 0.0290 *** |
| | (0.4110) | (0.0001) | (0.0001) |
| Poverty Rate | -0.0092 | 0.0074 | -0.0201 *** |
| | (0.5650) | (0.1220) | (0.0001) |
| **Economic** | | | |
| Percent Change in Number of Households 2000-2005 | -0.0563 *** | -0.0430 *** | -0.0386 *** |
| | (0.0001) | (0.0001) | (0.0001) |
| Percent of Houses Owner Occupied | -0.0164 | 0.0020 | -0.0039 * |
| | (0.1000) | (0.3390) | (0.0740) |
| Unemployment Rate | 0.0177 | 0.0150 | 0.0224 ** |
| | (0.5930) | (0.1240) | (0.0260) |
| Population: Employment Ratio | -1.1291 *** | -0.7886 *** | -1.0946 *** |
| | (0.0001) | (0.0001) | (0.0001) |
| Population: Proprietorship Ratio | 0.1057 *** | 0.0069 * | 0.0237 *** |
| | (0.0001) | (0.0620) | (0.0001) |
| **Organizations of Common Bond** | | | |
| Non-Agricultural Cooperatives Present | 0.5861 *** | 0.1908 *** | -0.2065 *** |
| | (0.0001) | (0.0001) | (0.0001) |
| Number of Civil-Social Organizations per 10K Population | 0.1502 *** | 0.0114 | -0.0048 |
| | (0.0001) | (0.4510) | (0.7510) |
| Number of Business Associations per 10K Population | -0.1402 ** | -0.2262 *** | -0.4194 *** |
| | (0.0040) | (0.0001) | (0.0001) |
| Number of Professional Associations per 10K Population | -0.0156 | 0.0973 ** | 0.2694 *** |
| | (0.8870) | (0.0200) | (0.0001) |
| Number of Labor Unions per 10K Population | 0.9006 *** | 0.3549 *** | 0.4996 *** |
| | (0.0001) | (0.0001) | (0.0001) |

Number in parentheses is the marginal significance value.
***: Significant at or above the 99.9% level.
**: Significant at the 95.0% level.
*: Significant at the 90.0% level.

Figure 1: Number of Credit Unions by Asset Peer Group

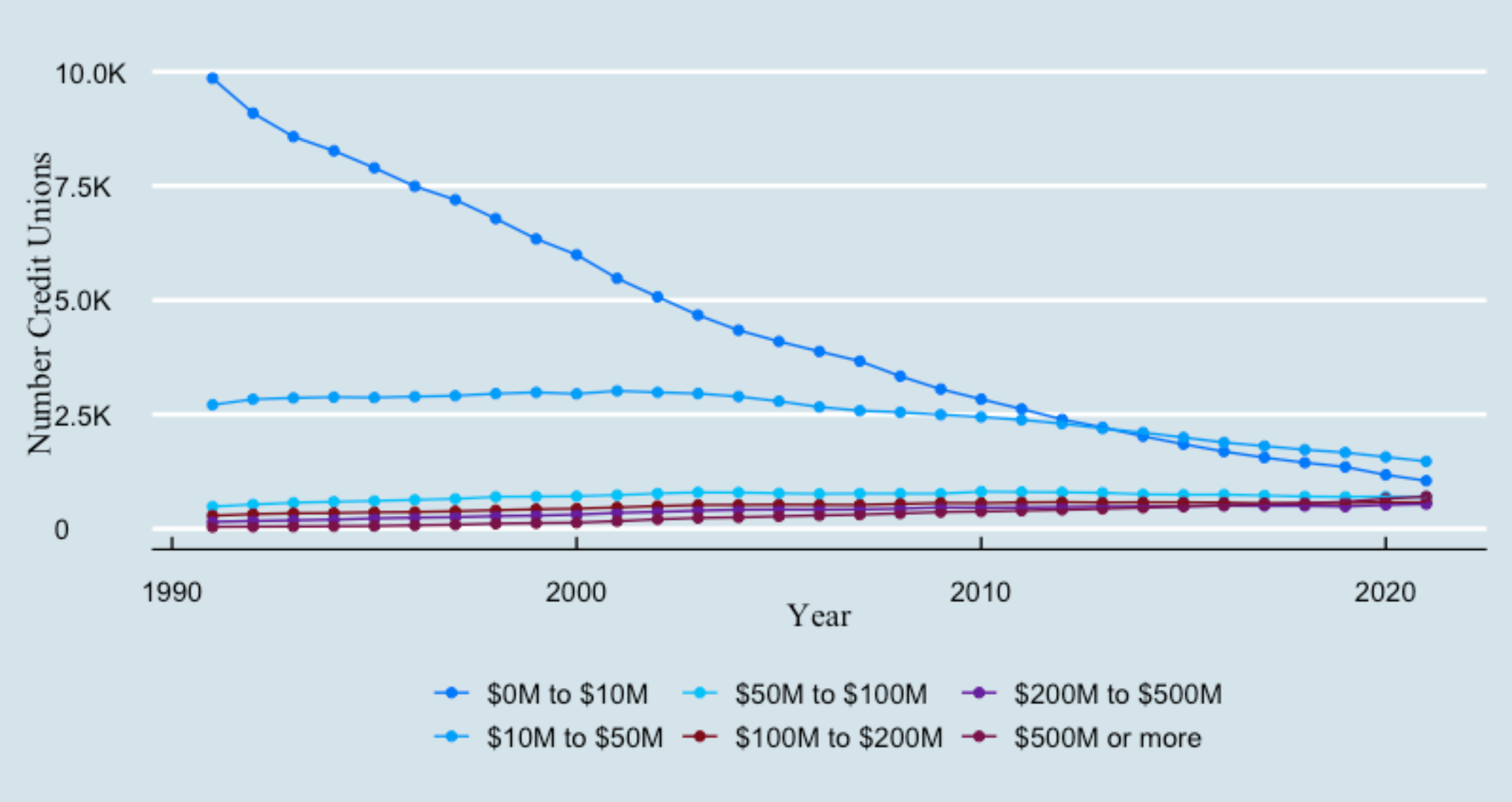

Source: NCUA call report data.

Figure 2: Number of Credit Union Members by Asset Peer Group

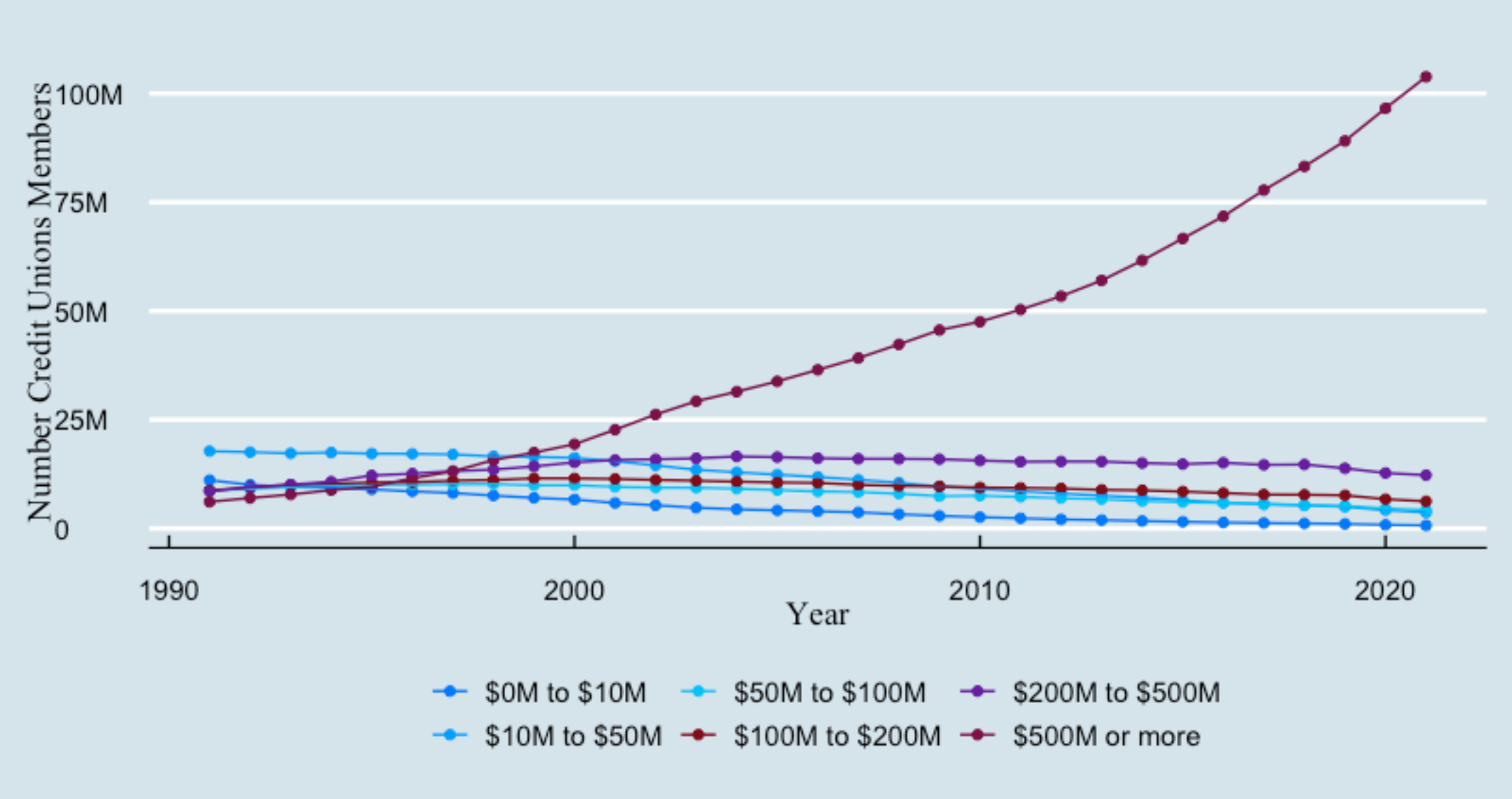

Source: NCUA call report data.

Figure 3: State of Financial Institutions by Type (2009).

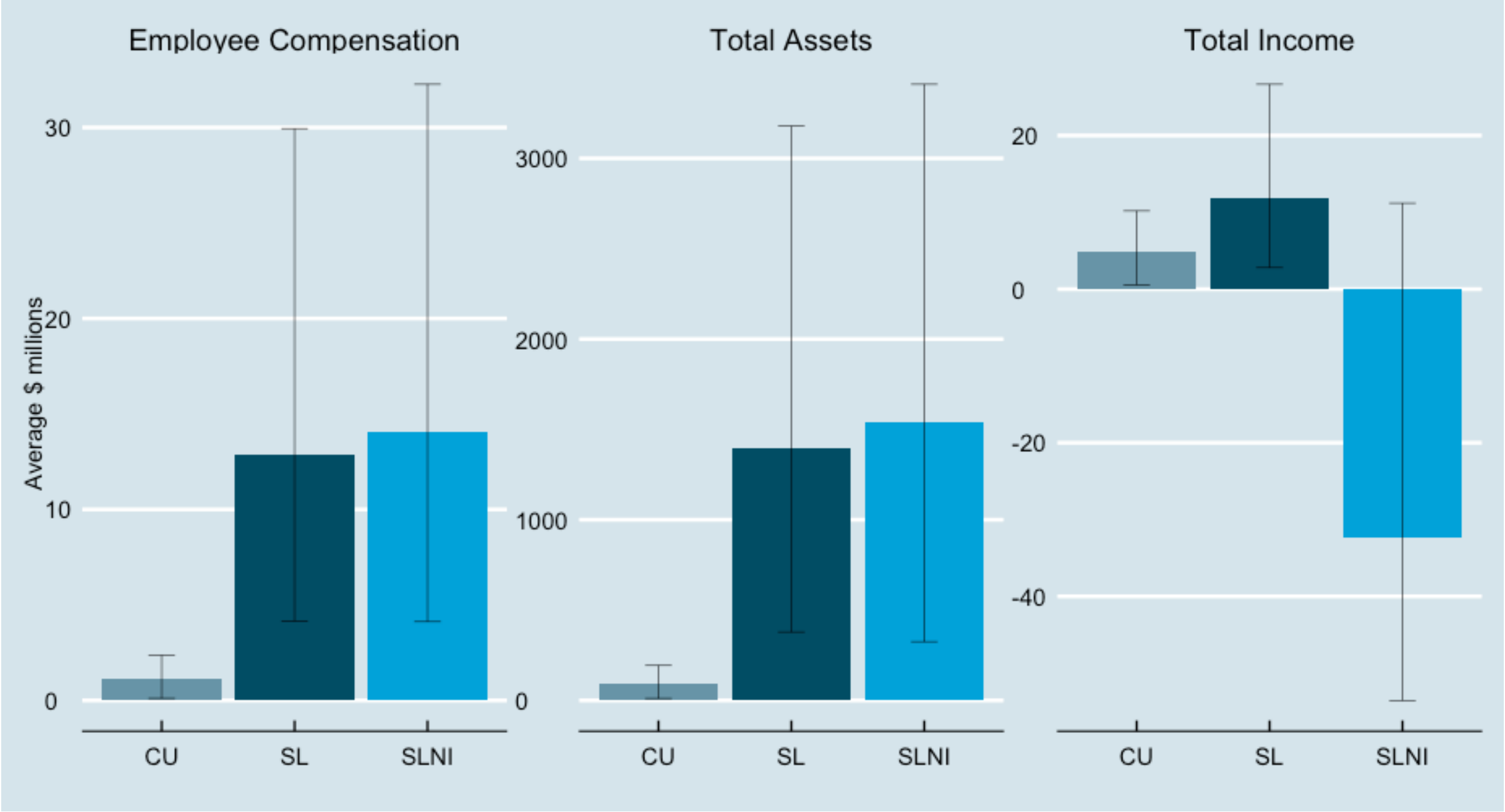

Source: NCUA call report data
CU = Credit Unions
SL = Thrifts, Savings and Loans
SLNI = Thrifts, Savings and Loans with Negative Total Income

Figure 4: Total Credit Union Assets by County in 2009.

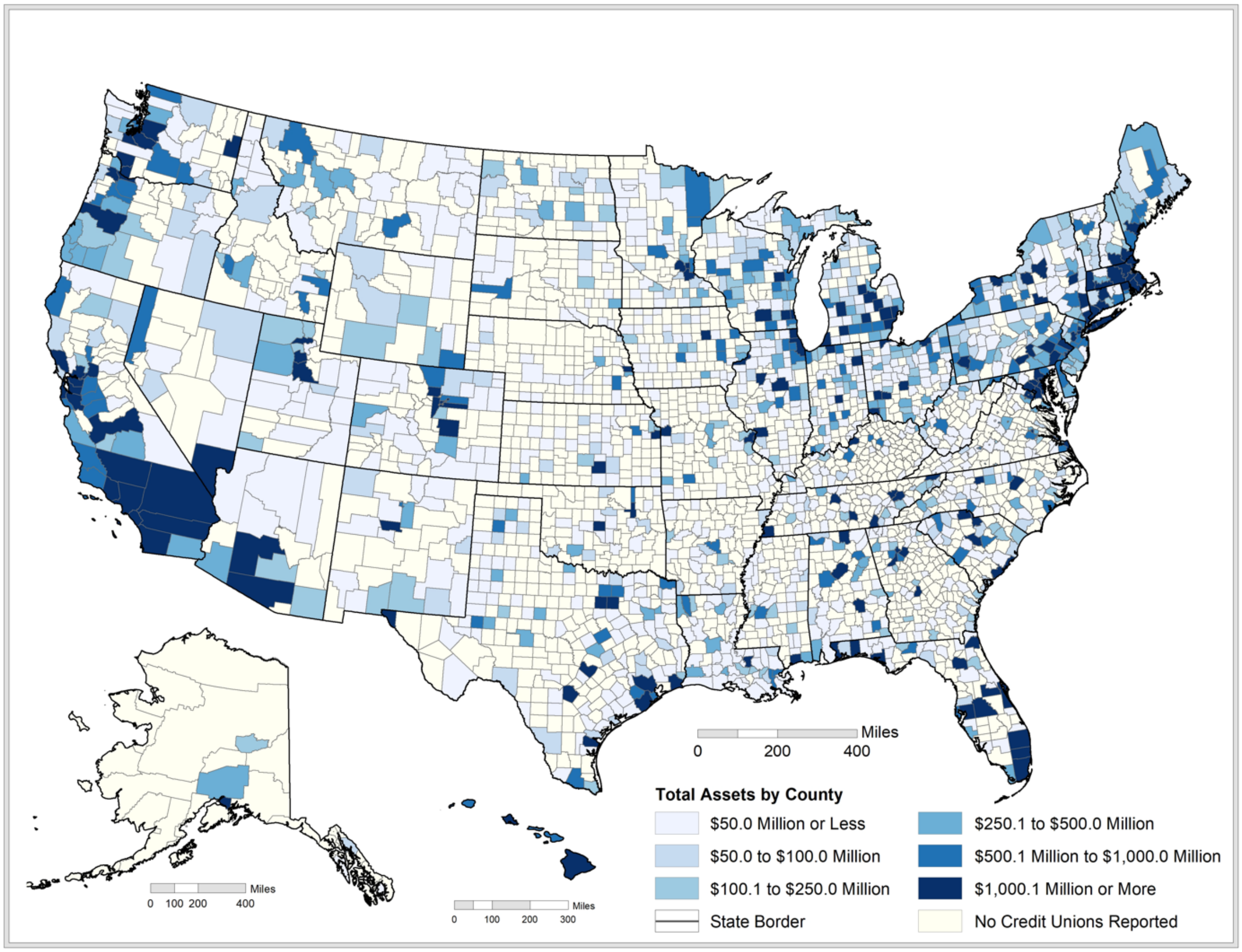

Source: NCUA call report data

Figure 5: Total Assets for Thrifts and Other Banks by County in 2009.

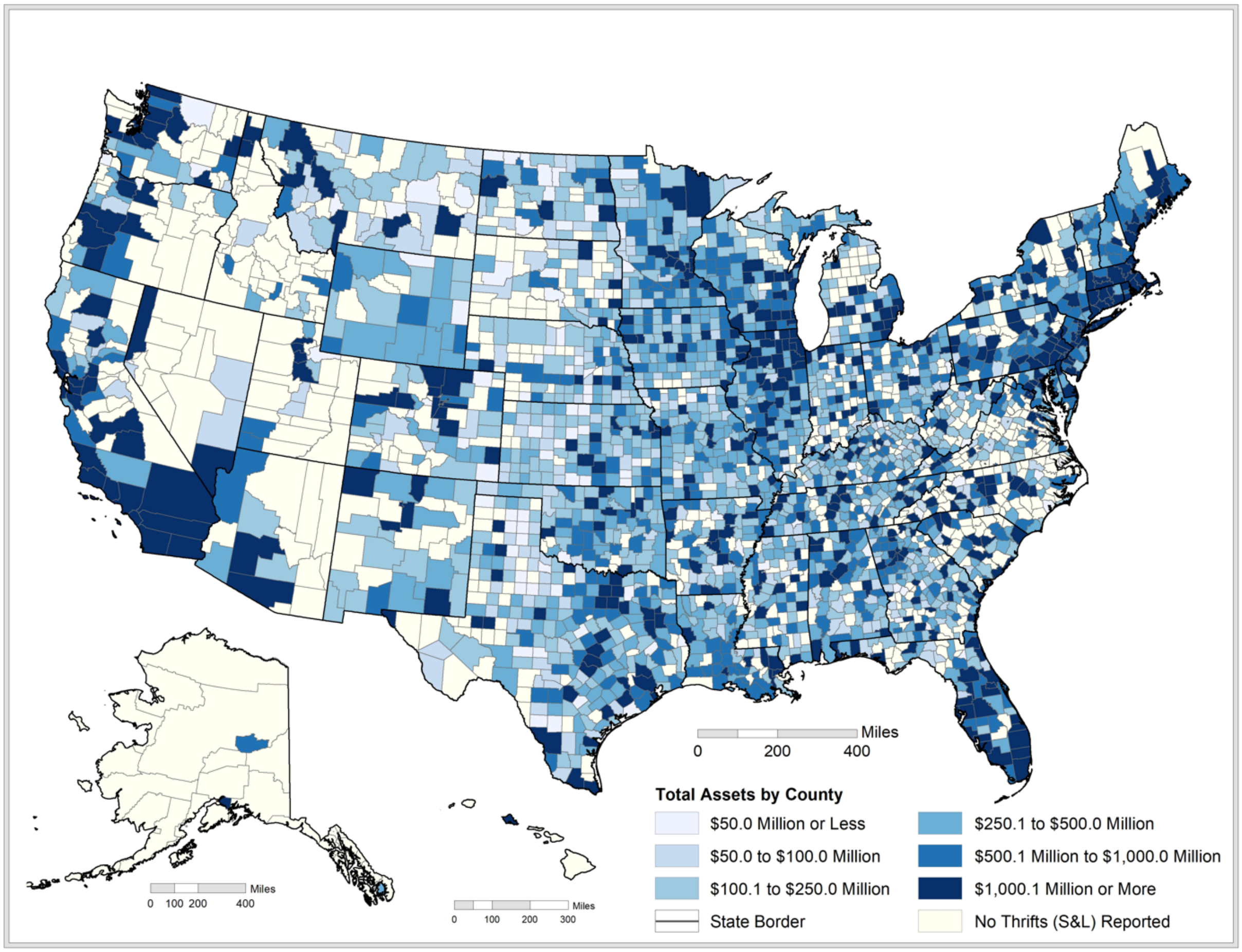

Source: NCUA call report data

Figure 6: Credit Union Location in 2009.

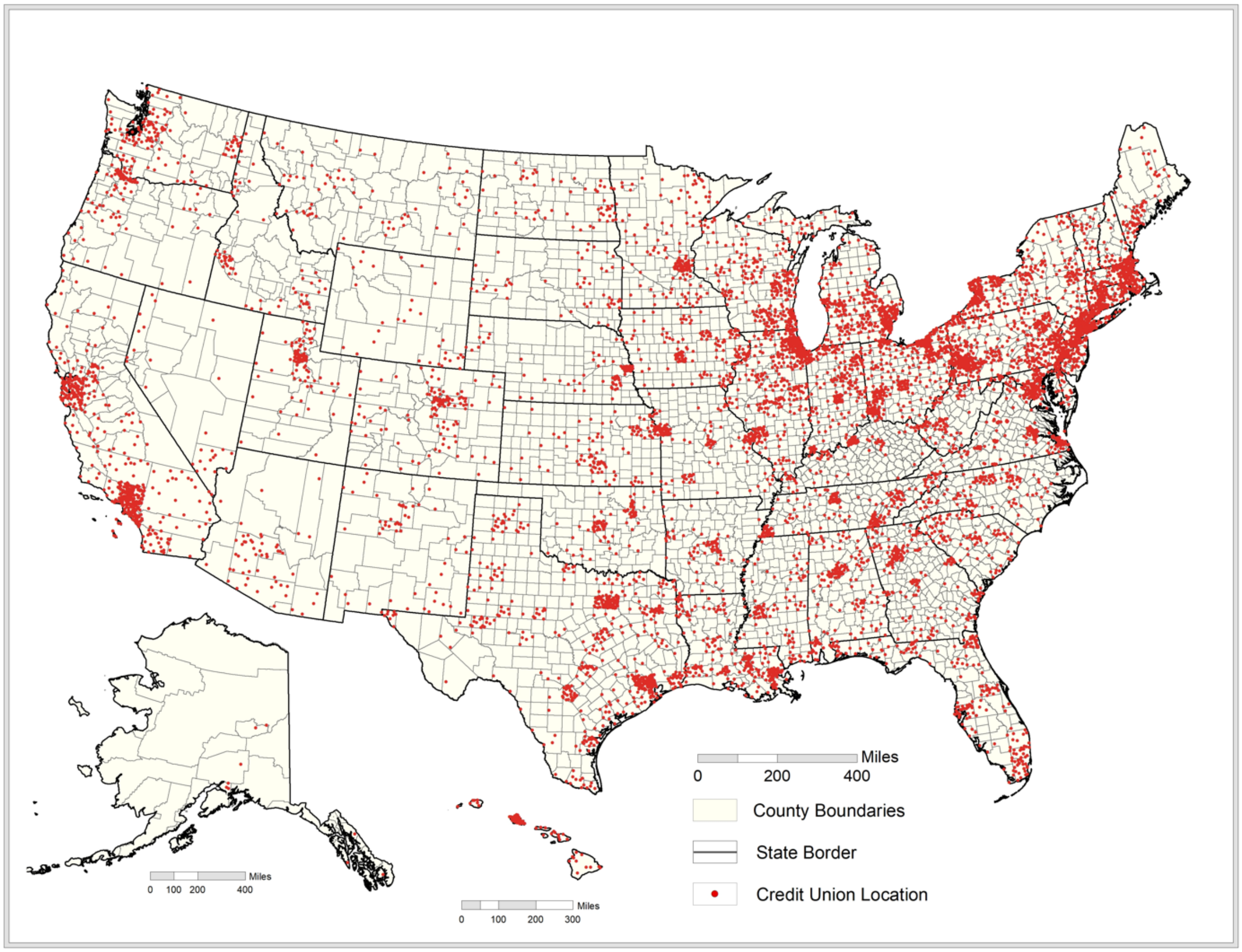

Source: NCUA call report data

Figure 7: Thrifts and Other Banks Locations in 2009.

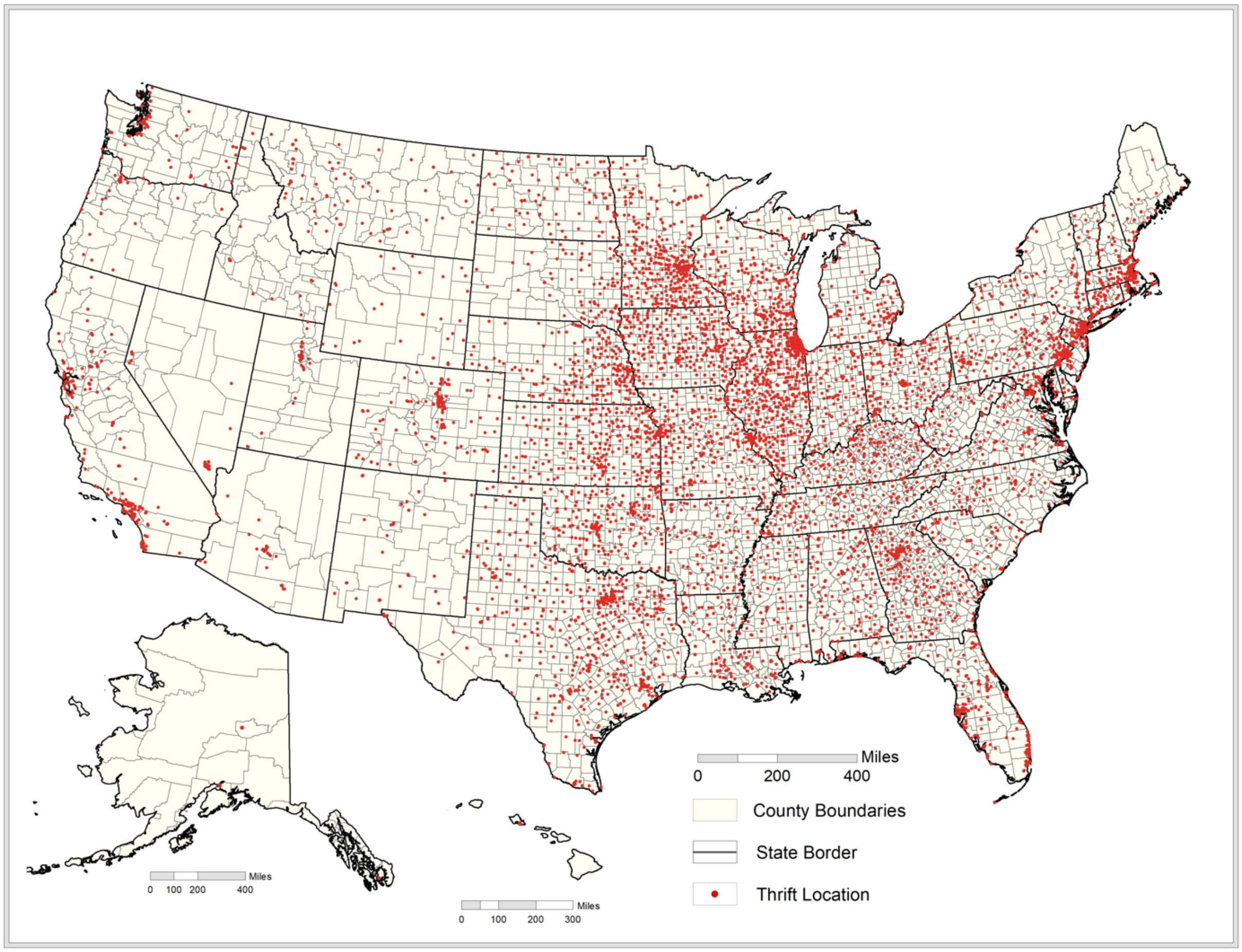

Source: NCUA call report data.

Figure 8: Credit Union Clustering: Number of Credit Unions.

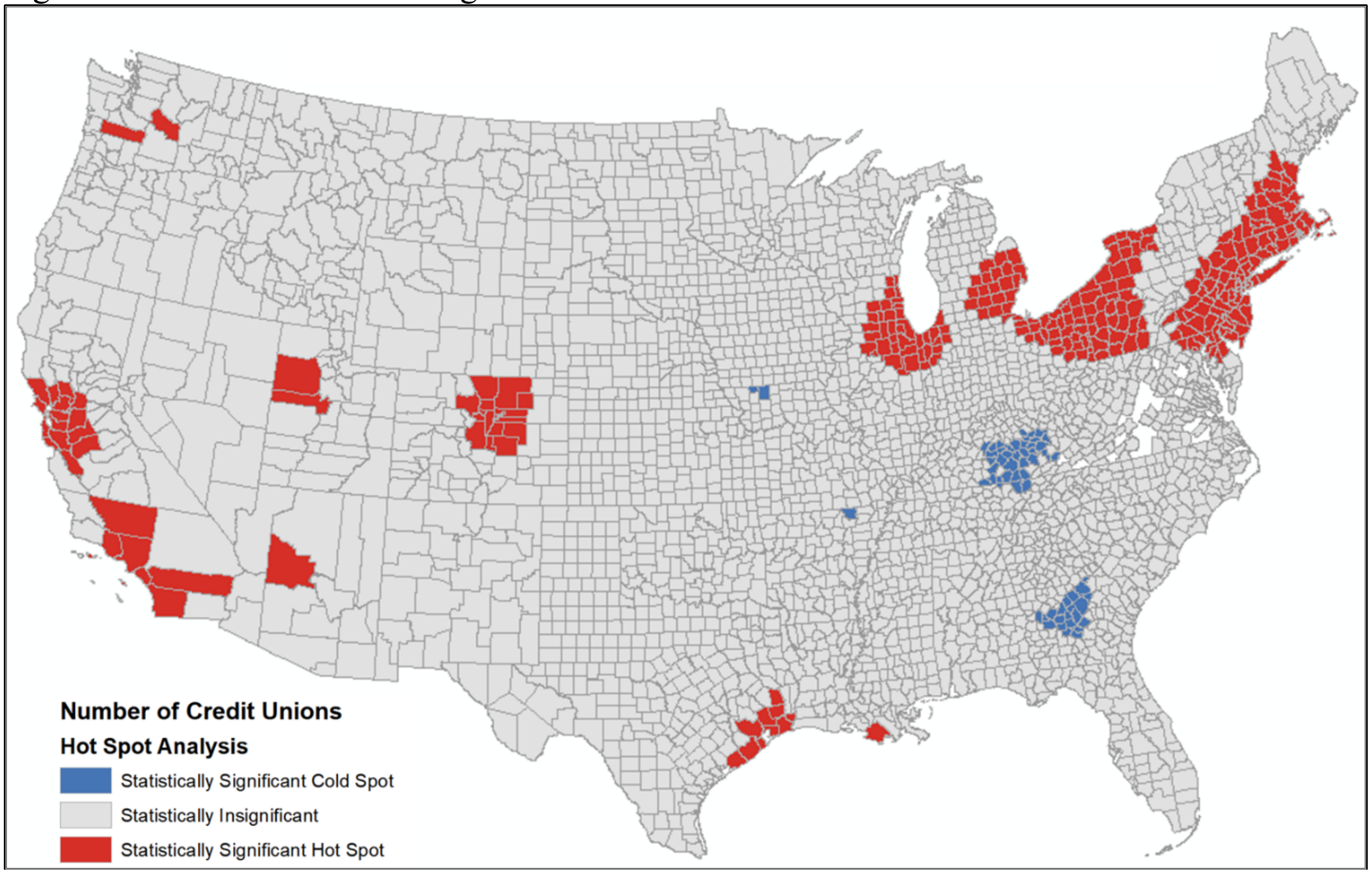

Source: NCUA call report data.